\documentclass[a4paper,11pt]{article}
\usepackage{jheppub} 
\usepackage{lineno}
\usepackage{hyperref}
\usepackage{latexsym,bm,amsmath,amssymb}
\usepackage{bbm}
\usepackage{physics}
\usepackage{color}
\usepackage{tikz}
\usepackage{tikz-feynman}
\usetikzlibrary{arrows}
\usepackage{graphicx}
\usepackage{multirow}
\usepackage{cancel}
\usepackage{diagbox}
\usepackage{makecell}
\usepackage{subcaption}
\usepackage{graphicx}
\newcommand{\br}{{\bm r}}
\newcommand{\bk}{{\bm k}}
\newcommand{\ba}{{\bm a}}

\newcommand{\bq}{{\bm q}}
\newcommand{\bi}{{\mathrm{i}}}
\newcommand{\me}{{\mathrm{e}}}
\newcommand{\calN}{{\mathcal{N}}}

\newcommand{\sgn}{{\mathrm{sgn}}}
\newcommand{\nn}{\nonumber \\}

\title{\boldmath Linear Resistivity from Spatially Random Interactions and the Uniqueness of Yukawa Coupling}

\author[a]{Sang-Jin Sin,~}
\author[a,b]{Yi-Li Wang}
\affiliation[a]{Department of Physics, Hanyang University,
	222 Wangsimni-ro, Seoul, 04763, Korea}
\affiliation[b]{Asia Pacific Center for Theoretical Physics, 77 Cheongam-ro, Pohang, 37673, Korea}

\emailAdd{sjsin@hanyang.ac.kr}
\emailAdd{yili.wang@apctp.org}

\abstract{Recent studies have shown that a spatially random Yukawa-type interaction between a Fermi surface and critical bosons can produce linear-in-temperature resistivity, the defining signature of strange metals. In this article, we systematically classify all scalar couplings of the form $(\psi^{\dagger}\psi)^n\phi^m$ in arbitrary dimensions to identify possible candidates for strange-metal behaviour within this disordered framework. We find that only spatially random Yukawa-type interaction in $(2+1)$ dimensions can yield linear resistivity. This indicates that linear resistivity is not a universal property of all spatially random scalar coupling, and strange-metal property replies on both dimensions and interaction type.}

\begin{document}
\maketitle
\flushbottom

\section{Introduction}
Strange metals have been one of the central topics in modern physics \cite{Lee2006,Anderson2017,Lee2018,Greene2020,Varma2020,Hartnoll2022,Phillips2022}. 
As the normal phase of superconductors, strange metals exhibit striking deviations from the Landau Fermi liquid theory. A defining feature of strange metals is a resistivity linear in temperature $T$, $\rho\sim \rho_0+AT$, 
where  $\rho_0$ is the residual resistivity and 
$A$ is a material-dependent coefficient. Despite extensive efforts over the past several decades, a systematic understanding of strange remains absent. A major obstacle is the lack of a theoretical framework accounting for the anomalous transport properties of strange metals, particularly the origin of linear resistivity.\\

Recently, Patel \textit{et al.} \cite{Patel2023} introduced a spatially random coupling between a Fermi Surface (FS) and a critical scalar boson, described by the interaction
\begin{eqnarray}\label{eqn:yukawa}
	S_{\text{int}}=\int d\tau d^2\br\sum_{i,j,l=1}^{N}\frac{g_{ijl}(\br)}{N}\psi^{\dagger}_i(\tau,\br)\psi_{j}(\tau,\br)\phi_l(\tau,\br).
\end{eqnarray}
The coupling constant $g_{ijl}(\br)$ follows a Gau\ss ian distribution with zero mean and variance $\langle g_{ijl}^*(\br)g_{i'j'l'}(\br')\rangle=g^2\delta(\br-\br')\delta_{ii',jj',ll'}$. 
At large-$N$ limit, this quenched disorder leads to linear resistivity at low temperatures.
It was pointed out in ref.\cite{Sin2025}  that in this space dependent random coupling's variance condition provides a wormhole picture in the field theory, which make it possible for far separated point to interact without distance dependent suppression.  
Subsequently, it was shown that the same strange metal transport property arises under an annealed average \cite{Sin2025}, and that  the equivalence between quenched and annealed averages can be viewed as a concrete realisation of the $ER=EPR$ conjecture~\cite{Maldacena2013}. 
 \\

Additionally, a similar fully randomised QED-like coupling between the critical Fermi surface and a vector boson can also reproduce such a linearity \cite{Wang:2024utm}, even in the presence of a magnetic field \cite{Wang2025}. These all-to-all spatially random interactions, inspired by the Sachdev-Ye-Kitaev (SYK) model \cite{Sachdev1993,Kitaev2015,Chowdhury2022}, are referred to as `SYK-rised' models in this article.
Although such couplings cannot capture other anomalies of strange metals, such as Hall angles \cite{Wang2025}, they currently offer the only known mechanism that produces linear resistivity in a controlled, analytically tractable setting.\\

SYK-rised models offer a promising route to understanding the nature of strange metals. This raises a natural question: \emph{what is most general types of random interactions between the FS and the critical boson that also give rise to linear resistivity?}
Beside higher-rank tensor couplings, one may consider interactions involving multiple fields.  Since the theory use the critical fluctuations, 
 we will investigate the SYK-rised interaction
 of the type  $(\psi^{\dagger}\psi)^n\phi^m$, 
 which does not generate mass scale, and investigate the possibility that leads to linear-$T$ resistivity across var$(2+1)$-dimensional Yukawa-type couplings are the only class that produces strange-metal behaviour. 
 \\

This article  is organised as follows. Section \ref{sec:model} develops a large-$N$ critical theory of spatial random couplings involving arbitrary numbers of fermions and bosons. The corresponding conductivity is calculated in Section \ref{sec:conductivity}. In Section \ref{sec:uniform}, we will show it is impossible to reproduce the linearity in spatially uniform models.  Concluding remarks are given in Section~\ref{sec:conclusion}.

\section{Generalised Spatially Random Coupling}\label{sec:model}
\subsection{$G-\Sigma$ formalism}
We begin by reviewing the model introduced in ref.\cite{Patel2023}, which provides the first theoretical realisation of linear-$T$ resistivity. The fermionic field $\psi_i$ and scalar field $\phi_i$ are described by standard kinetic terms,
\begin{eqnarray}
	S_0 &=& \int d\tau\sum_{\bk}\sum_{i=1}^N\psi^\dagger_{i\bk}(\tau)\left[\partial_\tau +\varepsilon (\bk)-\mu \right]\psi_{i\bk}(\tau) 
	+\int d\tau d^2\br \psi^\dagger_{i}(\tau,\br)v_{ij}(\br)\psi_j(\tau,\br)\nn
	&&+\frac{1}{2}\int d\tau \sum_{\bq}\sum_{i=1}^N \phi_{i\bq}(\tau)\left[-\partial_\tau^2 + \bq^2 +m_b^2\right]\phi_{i,-\bq}(\tau),
\end{eqnarray}
where $i=1,...,N$ denotes the flavour index. The potential disorder $v_{ij}$ is introduced with Gaus\ss ian statistics,
\begin{eqnarray}
	\langle v_{ij} (\br)\rangle = 0 \,, \quad \langle v^\ast_{ij} (\br) v_{lm} (\br')\rangle = v^2 \, \delta(\br-\br') \delta_{il} \delta_{jm}.
\end{eqnarray}
In this model, the key ingredient of the linearity is a quenched disorder interaction between electrons and bosons,
\begin{eqnarray}
	S_{\text{int}}=\int d\tau d^2\br \sum_{i,j,l=1}^N \frac{g_{ijl}(\br)}{N} \psi^\dagger_{i}(\br)\psi_{j}(\br)\phi_l(\br,\tau),
\end{eqnarray}
with the coupling constants satisfying
\begin{eqnarray}\label{eqn:int_0}
	\langle g_{jil}(\br)\rangle=0,\quad 
	\langle g^*_{ijl}(\br)g_{i'j'l'}(\br')\rangle=g^2\delta_{ii'jj'll'}\delta(\br-\br').
\end{eqnarray}
We refer to the interaction \eqref{eqn:int_0} as an `SYK-rised' Yukawa coupling,  as it describes a spatially random all-to-all interaction analogous to the SYK model.  It has been shown that in the large-$N$ limit, this interaction yields linear resistivity at low temperatures. Moreover, such SYK-rised mechanism remains effective when extended to vector bosons \cite{Wang:2024utm}. It is our interest to figure out whether this is the \emph{unique} scalar coupling that leads to linear resistivity, and whether such behaviour persists in higher-dimensional systems.\\

A straightforward generalisation of interaction \eqref{eqn:int_0} is to increase the number of fields and to consider an arbitrary dimension $d\geq 2$. Specifically, we consider an interaction fo the form
\begin{eqnarray}\label{eqn:multiple}
  S_{g}
  &=&\frac{g_{\{i\}\{j\}\{l\}}(\br)}{N^{(2n+m-1)/2}} \int d\tau d^d r \sum_{\{i\},\{j\},\{l\}}
  \psi^\dagger_{i_1,\bk}(\br,\tau)...\psi^\dagger_{i_n,\bk}(\br,\tau)\nn
  &\times&\psi_{j_1,\bk+\bq}(\br,\tau)...\psi_{j_n,\bk+\bq}(\br,\tau)\phi_{l_1,\bq}(\br,\tau)...\phi_{l_m,\bq}(\br,\tau).
  \label{eq:action}
\end{eqnarray}
The interaction involves $2n$ fermionic fields and $m$ bosonic fields, whilst the coupling constant $g_{\{i\}\{j\}\{l\}}\equiv g_{i_1...i_nj1...j_nl_1...l_m}$ obeys again Gau\ss ian distribution,  
\begin{eqnarray}
	&&\langle g_{\{i\}\{j\}\{l\}}(\br)\rangle=0,\\
	&&\langle g_{\{i\}\{j\}\{l\}}(\br)g^*_{\{i'\}\{j'\}\{l'\}}(\br')\rangle=g^2\delta(\br-\br')\delta_{\{i\}\{i'\},\{j\}\{j'\},\{l\}\{l'\}},
\end{eqnarray}
with multi-index delta functions defined by
\begin{eqnarray}
	\delta_{\{i\}\{i'\}}\equiv\delta_{i_1i_1'}\delta_{i_2i_2'}...\delta_{i_ni_n'}.
\end{eqnarray}
 Instead of working in $(2+1)$ dimensions, we consider general $(d+1)$D systems to find the effect of dimensionality. \\

We now derive the $G-\Sigma$ action of theory \eqref{eq:action}. We define bilocal variables
\begin{eqnarray}
	&&G(x_1,x_2)\equiv -\frac{1}{N}\sum_{i=1}^N\psi_{i}(x_1)\psi_{i}^{\dagger}(x_2),\\
	&&D(x_1,x_2)\equiv\frac{1}{N}\sum_{i=1}^N \phi_i(x_1)\phi_i(x_2),
\end{eqnarray}
and impose them via Lagrange multipliers as follows,
\begin{eqnarray}
	S_{L}&=&-N\int d\tau d\tau'\sum_k\Sigma(\bk,\tau'-\tau)\left[G(\bk,\tau-\tau')+\frac{1}{N}\sum_{i=1}^N\psi_{i,\bk}(\tau)\psi^\dagger_{i,\bk}(\tau')\right] \nn
	&&+\frac{N}{2}\int d\tau d\tau'\sum_q\Pi(\bq,\tau'-\tau)\left[D(\bq,\tau-\tau')-\frac{1}{N}\sum_{i=1}^N\phi_{i,\bq}(\tau)\phi_{i,-\bq}(\tau')\right].
\end{eqnarray}
Following the analysis in Ref.\cite{Esterlis2021}, we neglect the replica off-diagonal components of $G(x_1,x_2)$ and $D(x_1,x_2)$ in large $N$. As emphasised in Refs.\cite{Cardy1996,Esterlis2021}, the universal properties of critical theories are expected to be insensitive to the microscopic details of the interaction. We therefore assume replica symmetry from the outset.\\

Then we perform the disorder average using the replica trick. The partition function is given by
\begin{eqnarray}
	\mathcal{Z}\equiv\int\mathcal{D}[\psi,\psi^{\dagger}]\mathcal{D}[\phi]e^{-S_0-S_g-S_{\text{L}}},
\end{eqnarray}
and the resulting Gau\ss ian integrals can be evaluated as
\begin{eqnarray}
	&&\int \mathcal{D}[\psi,\psi^{\dagger}]e^{-\psi^{\dagger} \mathbf{A}\psi}=\det \mathbf{A},\\
	&&\int \mathcal{D}[\phi]e^{-\phi \mathbf{A}\phi/2}=\det \mathbf{A}^{-1/2}.
\end{eqnarray}
Applying these to the replicated action yields an effective action
\begin{eqnarray}\label{eqn:GS}
	\frac{S[G,\Sigma;D,\Pi]}{N}&=&-\ln\det(\partial_\tau +\varepsilon (\bk)\delta(x-x')-\mu+\Sigma)\nn
	&&+\frac{1}{2} \ln\det\left(\left(-\partial_\tau^2+K\bq^2+m_b^2\right)\delta(x-x')-\Pi\right)  \nn
	&&-\Tr\left(\Sigma\cdot G\right)+\frac{1}{2}\Tr\left(\Pi\cdot D\right)+\frac{v^2}{2}\Tr \left((G\bar{\delta})\cdot G\right)\nn
	&&+(-1)^{n^2+1}\frac{g^2}{2}\Tr\left((G^nD^m\bar{\delta})\cdot (G)^n \right),
\end{eqnarray}
where $\bar{\delta}$ is a Dirac delta over space, $\delta(\br-\br')$, and the trace notation is defined by \cite{Gu2020}
\begin{equation}\label{}
	\Tr(f\cdot g)\equiv f^T g\equiv\int d x_1 d x_2 f(x_2,x_1)g(x_1,x_2)\,.
\end{equation}
Assuming replica symmetry, we drop replica indices without loss of generality.

\paragraph{comments} In this article, each choice on $\{m,n,d\}$ defines a distinct and independent system. In particular, we do not study a tower of operators with different values of $\{m,n,d\}$ simultaneously. Our primary goal is to identify every choice of $\{m,n,d\}$ that can give rise to linear resistivity. Although this is represented by a single computation in the presentation below, our actual procedure is as follows:
\begin{itemize}
	\item[(1)] Fix $m$, $n$, and $d$;
	\item[(2)] Compute the corresponding scaling behaviour self-energies and resistivity, while temporarily ignoring details such as the explicit coefficient and potential divergences;
	\item[(3)] Only if a given choice of $\{m,n,d\}$ exhibits linear resistivity do we perform a more careful analysis, including the resolution of divergences;
	\item[(4)] Identify viable candidate theories of strange metals.
\end{itemize}
As each choice of $\{m,n,d\}$ defines an independent system, the self-energies and transport properties associated with other values of $\{m,n,d\}$ are irrelevant to the system under consideration.

\subsection{Large-$N$ Critical Theory}
The large-$N$ limit is governed by the saddle point of the $G-\Sigma$ action \eqref{eqn:GS}. Varying the action yields
\begin{eqnarray}\label{eq:deltaSstar}
	&&0=\frac{\delta S}{N}\\
	&&=\Tr\left(\delta\Sigma\cdot(G_*[\Sigma]-G)+\delta G\cdot(\Sigma_*[G]-\Sigma)+\frac{1}{2}\delta\Pi\cdot(D-D_*[\Pi])+\frac{1}{2}\delta D\cdot (\Pi-\Pi_*[D])\right)\nonumber.
\end{eqnarray}
Using the identity
\begin{equation}
	\ln \det \mathbf{M}=\Tr \ln \mathbf{M},  \qquad
	\delta\left[\ln\det\mathbf{M}\right]=\Tr \left[\mathbf{M}^{-1}\delta \mathbf{M}\right],
\end{equation}
we obtain the saddle-point equations (Schwinger-Dyson equations),
\begin{eqnarray}
	G_*[\Sigma](x_1,x_2) &=& (-\partial_\tau-\varepsilon(\bk)+\mu-\Sigma)^{-1}(x_1,x_2)\,,\label{eq:Gstar} \\
    \Sigma_*[G](x_1,x_2) &=&(-1)^{n^2+1} ng^2G^{2n-1}(x_1,x_2)D^m(x_1,x_2)\bar{\delta}+v^2G(x_1,x_2)\bar{\delta},\label{eq:Sigmastar} \\
	D_*[\Pi](x_1,x_2) &=& (-\partial_\tau^2+K\bq^2+m_b^2-\Pi)^{-1}(x_1,x_2),\label{eq:Dstar}\\
	\Pi_*[D](x_1,x_2) &=& mg^2 G^{2n}(x_1,x_2)D^{m-1}(x_1,x_2)\bar{\delta}.\label{eq:Pistar}
\end{eqnarray}
Setting $n=m=1$ recovers the saddle-point structure of the SYK-rised Yukawa model discussed in Refs.\cite{Patel2023,Esterlis2021,Guo2022}. We adopt the conventional quadratic dispersion $\varepsilon(\bk)=\bk^2/(2m)$, and fix units such that the boson velocity $\sqrt{K}=1$.\\

The next step is to solve these Schwinger-Dyson equations at zero temperature. To simplify the calculation, we define $\xi_{\bk}\equiv\varepsilon(\bk)-\mu$.
The contribution from potential disorder $v_{ij}(\br)$ is
\begin{eqnarray}\label{eqn:sigmav}
	\Sigma_{v}(\Omega_m)&=&v^2\int\frac{d^d\bk}{(2\pi)^2}G(i\Omega_m,\bk)\nn
	&=&v^2\calN\int d\xi_{\bk}\frac{1}{i\Omega_m-\xi_{\bk}-\Sigma(i\Omega_m,\bk)}\nn
	&=&-i \frac{\Gamma}{2}\sgn(\Omega_m),
\end{eqnarray}
where $\mathcal{N}$ is the fermion density of states \emph{at the Fermi level} in $d$-space, and  $\Gamma=2\pi v^2 \calN$ denotes the disorder scattering rate \cite{Coleman2019}. In the last step, we have used the fact that $\sgn(\omega_n)=-\sgn(\Im{\Sigma(\bi\omega_n)})$.\\

We now assume that impurity scattering dominates over interactions, \textit{i.e.} $|\Sigma_{v}|\gg|\Sigma_g|$, so that the electron propagator is well approximated by potential disorder alone. At low frequencies,
\begin{eqnarray}\label{eqn:g1st}
	G(\bi\omega,\bk)&\simeq&\frac{1}{\bi\sgn(\omega)\Gamma/2-\bk^2/(2m)+\mu},
\end{eqnarray}
where the corrections from fermion-boson coupling are suppressed at leading order.\\

At critical point, the boson mass satisfies $m_b^2-\Pi(0,0)=0$ \cite{Esterlis2021}. The full bosonic propagator can be expressed as
\begin{eqnarray}
	D(\bi\Omega_m,\bq)
	&=&\frac{1}{\Omega_m^2+\bq^2-(\Pi-\Pi(\Omega_m=0))}.
\end{eqnarray}
It remains to evaluate $\Pi(\bi\Omega)-\Pi(0,0)$, which encodes the leading dynamical correction to the critical bosonic propagator.\\  

\begin{figure}
	\centering
	\begin{subfigure}[c]{0.45\columnwidth}
		\centering
		\includegraphics[width=0.7\textwidth]{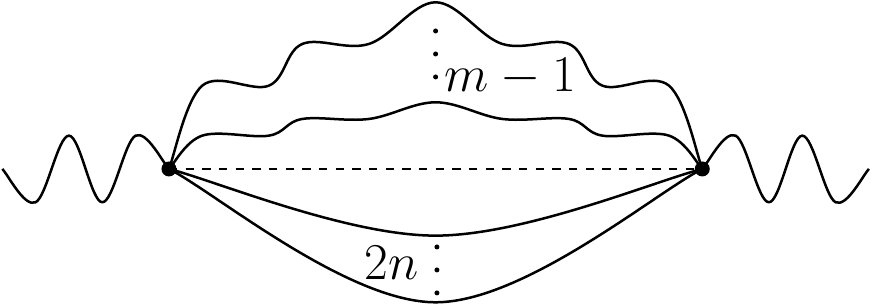}
		\caption{\label{fig:bse}Bosonic self-energy}
	\end{subfigure}
	\begin{subfigure}[c]{0.45\columnwidth}
		\centering
		\includegraphics[width=0.7\textwidth]{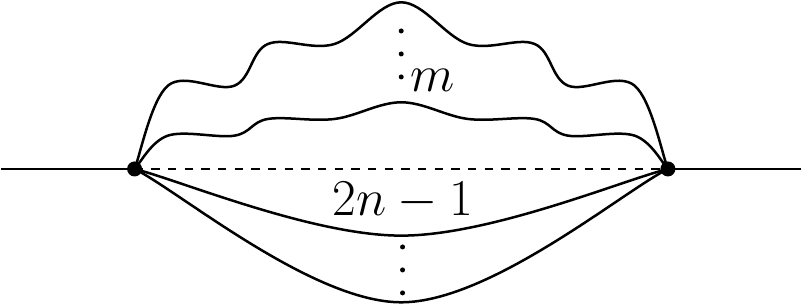}
		\caption{\label{fig:fse}Fermionic self-energy}
	\end{subfigure}
	\caption{\label{fig:se} Two self-energies, $\Pi$ and $\Sigma$, share the same structure.
	}
\end{figure}

We now turn to the evaluation of the self-energies. 
The boson self-energy \eqref{eq:Pistar} is graphically represented by Fig.\ref{fig:bse}, where solid lines denote fermion propagators, wavy lines denote bosons, and dashed lines indicate disorder averaging. The fermionic self-energy \eqref{eq:Sigmastar} is illustrated in Fig.\ref{fig:fse}.
Since bosonic self-energy and bosonic self-energy are structurally similar, we do not need to compute them separately. 
Instead, we can analyse both simultaneously by studying the frequency dependence of a generic diagram containing $\alpha$ fermion lines and $\beta$ boson lines. \\


The fermionic Green's function is approximately given by eqn.\eqref{eqn:g1st}, but bosonic propagator remains undetermined. Though generic diagrams with $\beta\geq1$ appear intractable, they can still be analysed by assuming  a scaling form for the bosonic self-energy, 
\begin{eqnarray}
	\Pi(\bi\Omega_m)-\Pi(0)\sim -c_B |\Omega_m|^{\eta'},
\end{eqnarray}
where $\eta'$ will be fixed self-consistently at the end of the calculation. At low frequencies, if $\eta< 2$, the bosonic propagator reads 
\begin{eqnarray}
	D(\bi\Omega_m,\bq)\simeq\frac{1}{\bq^2+c_B|\Omega_m|^{\eta'}}.
\end{eqnarray}
Otherwise, the frequency dependence from self-energies is subleading, and
\begin{eqnarray}
	D(\bi\Omega_m,\bq)\simeq\frac{1}{\Omega_m^2+\bq^2}.
\end{eqnarray}
In either case, the propagator may be written uniformly as
\begin{eqnarray}
	D(\bi\Omega_m,\bq)\simeq\frac{1}{\bq^2+c_B|\Omega_m|^{\eta}},
\end{eqnarray}
with the requirement that $\eta\leq2$.\\

Therefore, both fermion and boson self-energies take the form 
\begin{eqnarray}\label{eqn:int}
	\mathcal{I}^d_{\alpha,\beta}(\bi x)
	&\equiv&\int_{-\infty}^{\infty}\left(\prod_{i=1}^{\alpha}\frac{d\omega_i}{2\pi}d\xi_{\bk_i}\frac{1}{\bi\sgn(\omega_i)\Gamma/2-\xi_{\bk_i}}\right)\nn
	&&\times\left(\int_{-\infty}^{\infty}\prod_{j=1}^{\beta-1}\frac{d\Omega_j}{2\pi}\int_{0}^{\infty}\frac{d^d\bq_j}{(2\pi)^2}
	\frac{1}{c_B|\Omega_j|^\eta+\bq_j^2}\right)
	\int_0^{\infty}\frac{d^d\bq_\beta}{(2\pi)^2}\frac{1}{c_B|x+\sum_{j=1}^{\beta-1}\Omega_j+\sum_{i=1}^{\alpha}\omega_i|^\eta+\bq_\beta^2}.\nn
\end{eqnarray}
To evaluate the integral \eqref{eqn:int}, we should first notice that the dimension plays an important role. When $d=2$, one finds
\begin{eqnarray}
	\int_0^\infty d^2 \bq\frac{1}{\bq^2+A}=\frac{1}{2}\ln(\frac{\Lambda_q^2}{A}),
\end{eqnarray}
where $\Lambda_q$ is the UV cutoff on $\bq$. For $d\geq 3$, the integral yields
\begin{eqnarray}\label{eqn:qint}
	\int_0^\infty d^d \bq\frac{1}{\bq^2+A}=\sum_{i=0}^{\lceil (d-2)/2 \rceil-1}c_i\Lambda_q^{d-2-2i}A^i + A^{\frac{d-2}{2}},
\end{eqnarray}
with $c_i$ the coefficient and $\lceil ... \rceil$ the ceiling function.\\

Since cases where $d=2$ and $d>2$ are fundamentally different, we will evaluate them separately. Let us first compute the full integral \eqref{eqn:int} in $(2+1)$ dimensions:
\begin{eqnarray}
	\mathcal{I}^{d=2}_{\alpha,\beta}(\bi x)
	&\equiv&\int_{-\infty}^{\infty}\left(\prod_{i=1}^{\alpha}\frac{d\omega_i}{2\pi}d\xi_{\bk_i}\frac{1}{\bi\sgn(\omega_i)\Gamma/2-\xi_{\bk_i}}\right)\nn
	&&\times\left(\prod_{j=1}^{\beta-1}\int_{-\infty}^{\infty}\frac{d\Omega_j}{2\pi}\int_{0}^{\infty}\frac{d^2\bq_j}{(2\pi)^2}
	\frac{1}{|\Omega_j|^\eta+\bq_j^2}\right)
	\int_0^{\infty}\frac{d^2\bq_\beta}{(2\pi)^2}\frac{1}{|x+\sum_{j=1}^{\beta-1}\Omega_j+\sum_{i=1}^{2n}\omega_i|^\eta+\bq_\beta^2}\nn
	&\sim&(-\bi)^{\alpha}\int_{-\infty}^{\infty}\left(\prod_{i=1}^{\alpha}d\omega_i\sgn(\omega_i)\right)\left(\prod_{j=1}^{\beta-1}\int_{-\infty}^{\infty}d\Omega_j
	\ln(\frac{\Lambda_q^2}{|\Omega_j|^\eta})\right)\ln(\frac{\Lambda_q^2}{|x+\sum_{j=1}^{\beta-1}\Omega_j+\sum_{i=1}^{2n}\omega_i|^\eta})\nn
	&\sim&(-\bi)^{\alpha}\left(\prod_{j=1}^{\beta-1}\int_{-\infty}^{\infty}d\Omega_j
	\ln(\frac{\Lambda_q^2}{|\Omega_j|^\eta})\right)\left(\left(x+\sum_{j=1}^{\beta-1}\Omega_j\right)^{\alpha}\left(C_1\ln\left|x+\sum_{j=1}^{\beta-1}\Omega_j\right|+C_2\right)\right).
\end{eqnarray}
Changing the variables $\{\Omega_i\}$ into $\{u_i\}$, such that
\begin{eqnarray}
	\Omega_j\equiv x\cdot u_j \Leftrightarrow d\Omega_j= x du_j,
\end{eqnarray}
one finds
\begin{eqnarray}\label{eqn:2d}
	\mathcal{I}_{\alpha,\beta}^{d=2}(\bi x)&\sim&(-\bi)^{\alpha}x^{\alpha+\beta-1}\int \prod_{j=1}^{\beta-1}du_j\ln(x\cdot u_j)\left(1+\sum_{j=1}^{\beta-1}u_j\right)^{\alpha}
	\ln(x\left(1+\sum_{j=1}^{\beta-1}u_j\right))\nn
	&\sim&(-\bi)^{\alpha}x^{\alpha+\beta-1}\sum_{\iota=0}^{\beta}c_{\iota}(\ln x_i)^\iota.
\end{eqnarray}
The coefficient $c_i$ is fixed by explicit computation, but being $x$-independent, can remain unspecified in a qualitative analysis.\\

Applying eqn.\eqref{eqn:qint}, we can evaluate the scaling behaviour when $d\geq 3$, where
\begin{eqnarray}\label{eqn:dintegral1}
	\mathcal{I}^{d\geq3}_{\alpha,\beta}(\bi x)
	&\equiv&\int_{-\infty}^{\infty}\left(\prod_{i=1}^{\alpha}\frac{d\omega_i}{2\pi}d\xi_{\bk_i}\frac{1}{\bi\sgn(\omega_i)\Gamma/2-\xi_{\bk_i}}\right)\nn
	&&\times\left(\int_{-\infty}^{\infty}\prod_{j=1}^{\beta-1}\frac{d\Omega_j}{2\pi}\int_{0}^{\infty}\frac{d^d\bq_j}{(2\pi)^2}
	\frac{1}{c_B|\Omega_j|^\eta+\bq_j^2}\right)
	\int_0^{\infty}\frac{d^d\bq_\beta}{(2\pi)^2}\frac{1}{c_B|x+\sum_{j=1}^{\beta-1}\Omega_j+\sum_{i=1}^{2n}\omega_i|^\eta+\bq_\beta^2}\nn
	&\sim&(-\bi)^{\alpha}\int_{-\infty}^{\infty}\left(\prod_{i=1}^{\alpha}d\omega_i\sgn(\omega_i)\right)\left(\prod_{j=1}^{\beta-1}\int_{-\infty}^{\infty}d\Omega_j
	\sum_{l=0}^{\lceil (d-2)/2\rceil-1}c_l\Lambda_q^{d-2-2l}|\Omega_j|^{\eta \cdot l}+c_h\Omega_j^{\frac{(d-2)\eta}{2}}\right)\nn
	&&\times
	\left(\sum_{l=0}^{\lceil (d-2)/2\rceil-1}c_l'\Lambda_q^{d-2-2l}\left|x+\sum_{j=1}^{\beta-1}\Omega_j+\sum_{i=1}^{2n}\omega_i\right|^{\Theta(\beta)\eta \cdot l}+c_h'\left|x+\sum_{j=1}^{\beta-1}\Omega_j+\sum_{i=1}^{2n}\omega_i\right|^{\frac{(d-2)\eta}{2}}\right),\nn
\end{eqnarray}
where the Heaviside theta function $\Theta(\beta)$ indicates that the dependence on $\eta$ from $\int d^d\bq$ only appears if $\beta\geq 1$.
In the equation above, the polynomial $\sum_{l=0}^{d-2}c_l'\Lambda_q^{d-2-l}\left|x+\sum_{j=1}^{\beta-1}\Omega_j+\sum_{i=1}^{2n}\omega_i\right|^{\frac{\eta \cdot l}{2}}$ contains a term independent of $\omega_i$ when $l=0$, which is of order $\Lambda_q^{d-2}$. Since $\sgn(\omega_i)$ is an odd function of $\omega_i$, this term will vanish. Consequently, the integral becomes
\begin{eqnarray}
	\mathcal{I}^{d\geq3}_{\alpha,\beta}(\bi x)&\sim&
	(-\bi)^{\alpha}\int_{-\infty}^{\infty}\left(\prod_{i=1}^{\alpha}d\omega_i\sgn(\omega_i)\right)\left(\prod_{j=1}^{\beta-1}\int_{-\infty}^{\infty}d\Omega_j
	\sum_{l=0}^{\lceil (d-2)/2\rceil-1}c_l\Lambda_q^{d-2-2l}|\Omega_j|^{\eta \cdot l}+c_h\Omega_j^{\frac{(d-2)\eta}{2}}\right)\nn
	&&\times\left(\sum_{l=1}^{\lceil (d-2)/2\rceil-1}c_l'\Lambda_q^{d-2-2l}\left|x+\sum_{j=1}^{\beta-1}\Omega_j+\sum_{i=1}^{2n}\omega_i\right|^{\Theta(\beta)\eta \cdot l}+c_h'\left|x+\sum_{j=1}^{\beta-1}\Omega_j+\sum_{i=1}^{2n}\omega_i\right|^{\Theta(\beta)\frac{(d-2)\eta}{2}}\right)\nn
	&\sim&(-\bi)^{\alpha}
	\left(\prod_{j=1}^{\beta-1}\int_{-\infty}^{\infty}d\Omega_j
	\sum_{l=0}^{\lceil (d-2)/2\rceil-1}c_l\Lambda_q^{d-2-2l}|\Omega_j|^{\Theta(\beta)\eta \cdot l}+c_h\Omega_j^{\Theta(\beta)\frac{(d-2)\eta}{2}}\right)\\
	&&\times \left(\sum_{l=1}^{\lceil (d-2)/2\rceil-1}c_l'\Lambda_q^{d-2-2l}\left|\sum_{j=1}^{\beta-1}\Omega_j+x\right|^{\Theta(\beta)\eta \cdot l+\alpha}+c_h'\left|\sum_{j=1}^{\beta-1}\Omega_j+x\right|^{\Theta(\beta)\frac{(d-2)\eta}{2}+\alpha}\right)\nonumber
\end{eqnarray}
We again perform the change of variables $\Omega_j\equiv x\cdot u_j$ so that $d\Omega_j= x du_j$, so
\begin{eqnarray}
	\mathcal{I}^{d\geq3}_{\alpha,\beta}(\bi x)&\sim&
	(-\bi)^{\alpha}x^{\beta-1}
	\left(\prod_{j=1}^{\beta-1}\left(
	\sum_{l=0}^{\lceil (d-2)/2\rceil-1}c_l\Lambda_q^{d-2-l}|x_j|^{\eta \cdot l}+c_h x^{\Theta(\beta)\frac{(d-2)\eta}{2}}\right)\right)\nn
	&&\times
	\left(\sum_{l=1}^{\lceil (d-2)/2\rceil-1}c_l'\Lambda_q^{d-2-2l}\left|x\right|^{\Theta(\beta)\eta \cdot l+\alpha}+c_h'\left|x\right|^{\Theta(\beta)\frac{(d-2)\eta}{2}+\alpha}\right)\nn
	&=&\sum_l c^l x^{l},
\end{eqnarray}
where $l$ satisfies
\begin{eqnarray}
	\min\{l\}=	\left\{\begin{aligned}
		[\Theta(\beta)\eta/2+\alpha+\beta-1], & d=3\\
		[\Theta(\beta)\eta+\alpha+\beta-1], &d\geq 4
	\end{aligned}
	\right., \eta>0,
\end{eqnarray}
and
\begin{eqnarray}
	\min\{l\}=	\Theta(\beta)\beta\eta(d-2)/2+\alpha+\beta-1,
	\,\eta<0.
\end{eqnarray}
We merely focus on the scaling behaviour of $x$, so the integral $\int du$ is not considered.
\\

For bosonic self-energy, we identify $\alpha=2n$ and $\beta=m-1$; for fermionic self-energy, $\alpha=2n-1$ and $\beta=m$.\footnote{Our result also captures the feature of models with $m=1$.} We are interested in low frequencies, where $|x|\ll 1$. So the results can be divided into three categories.
\begin{itemize}
	\item If $d=2$, according to eqn.\eqref{eqn:2d}, the leading order of self-energies are
	\begin{eqnarray}
		\Pi(\bi\Omega)-\Pi(0)&\sim& -c_Bg^2\Omega^{2n+m-2}\sum_{\iota=0}^{m-1}c_{\iota}(\ln x_i)^\iota,\label{eqn:d2pi}\\
		\Sigma(\bi\omega)&\sim& -\bi c_Fg^2\omega^{2n+m-2}\sum_{\iota=0}^{m}c_{\iota}(\ln x_i)^\iota. \label{eqn:d2sigma}
	\end{eqnarray}
	\item For $d=3$, there are two cases. 
	\begin{itemize}
		\item If $\eta>0$, 
			\begin{eqnarray}
			\Pi(\bi\Omega)-\Pi(0)&\sim& -c_Bg^2\Omega^{\Theta(m-1)\frac{\eta}{2}+2n+m-2},\\
			\Sigma(\bi\omega)&\sim& -\bi c_Fg^2\omega^{\Theta(m)\frac{\eta}{2}+2n+m-2}. \label{eqn:d3sigma1}
		\end{eqnarray}
		\item If $\eta<0$,
		\begin{eqnarray}
			\Pi(\bi\Omega)-\Pi(0)&\sim& -c_Bg^2\Omega^{\Theta(m-1)\frac{\eta(m-1)}{2}+2n+m-2},\\
			\Sigma(\bi\omega)&\sim& -\bi c_Fg^2\omega^{\Theta(m)\frac{\eta m }{2}+2n+m-2}.\label{eqn:d3sigma2}
		\end{eqnarray}
	\end{itemize}
	\item For $d\geq 4$, there are two cases as well. 
	\begin{itemize}
		\item If $\eta>0$, 
		\begin{eqnarray}
			\Pi(\bi\Omega)-\Pi(0)&\sim& -c_Bg^2\Omega^{\Theta(m-1){\eta}+2n+m-2},\\
			\Sigma(\bi\omega)&\sim& -\bi c_Fg^2\omega^{\Theta(m){\eta}+2n+m-2}. \label{eqn:d4sigma1}
		\end{eqnarray}
		\item If $\eta<0$,
		\begin{eqnarray}
			\Pi(\bi\Omega)-\Pi(0)&\sim& -c_Bg^2\Omega^{\Theta(m-1)\frac{\eta(m-1)(d-2)}{2}+2n+m-2},\\
			\Sigma(\bi\omega)&\sim& -\bi c_Fg^2\omega^{\Theta(m)\frac{\eta m (d-2)}{2}+2n+m-2}.\label{eqn:d4sigma2}
		\end{eqnarray}
	\end{itemize}
\end{itemize}
Here, $c_B$ and $c_F$ constants not relevant for scaling analysis.\\

Before we move on, we should discuss the issue is the divergence of self-energies (and possibly the conductivity). We can see that eqn.\eqref{eqn:dintegral1} captures the scaling behaviour of $x$ (the frequencies), but the integral $\int du$, which determines the coefficient, is generally divergent. In Yukawa-type model \eqref{eqn:yukawa}, where $m=n=1$, the quantity $\Pi(\bi\Omega)-\Pi(0)$ is finite, which is an exception. Moreover, even when $m=n=1$, the electron self-energy $\Sigma$ is accompanied by a UV divergent term \cite{Patel2023}. This divergence only appears in the imaginary part of the conductivity, so it can be ignored. However, neither self-energies nor conductivity is necessarily finite when we consider general $m$ and $n$. The divergence due to integral $\int du$ implies that we need a cutoff on the frequencies $\Omega_j$, which requires careful treatment. For the time being, let us set aside this divergence, and focus only on the scaling behaviour first. We consider the regularisation only when the divergence appears in models yielding linear resistivity. \\

\section{Conductivity}\label{sec:conductivity}
\subsection{The unique approach to linearity}
Having obtained the propagators and self-energies, we now estimate the conductivity using the \emph{Kubo formula},
\begin{eqnarray}\label{eq:kubo}
	\sigma^{\mu\nu}(\bi\Omega_m)&=&
	-\frac{1}{\Omega_m}\left[\tilde{\Pi}^{\mu\nu}(\Omega)\right]_{\Omega=0}^{\Omega=\bi\Omega_m}
\end{eqnarray}
where $\tilde{\Pi}^{\mu\nu}$ denotes the current-current correlator, or equivalently, the polarisation bubble for the external electromagnetic field. Since we consider only electric fields in this work, $\tilde{\Pi}^{\mu\nu}$ is diagonal.\\ 
\begin{figure}
	\centering
	\begin{subfigure}[b]{0.4\columnwidth}
		\centering
		\includegraphics[width=\textwidth]{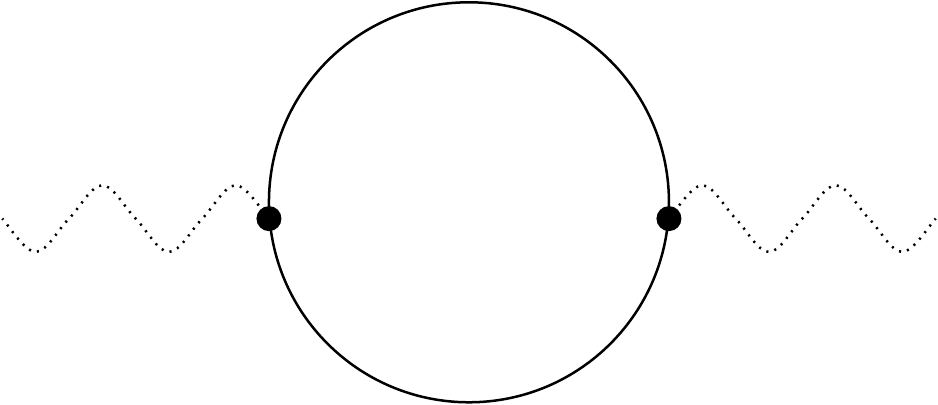}
		\caption{The simplest polarisation bubble.}
		\label{fig:jj}
	\end{subfigure}\hfil
	\begin{subfigure}[b]{0.4\columnwidth}
		\centering
		\includegraphics[width=\textwidth]{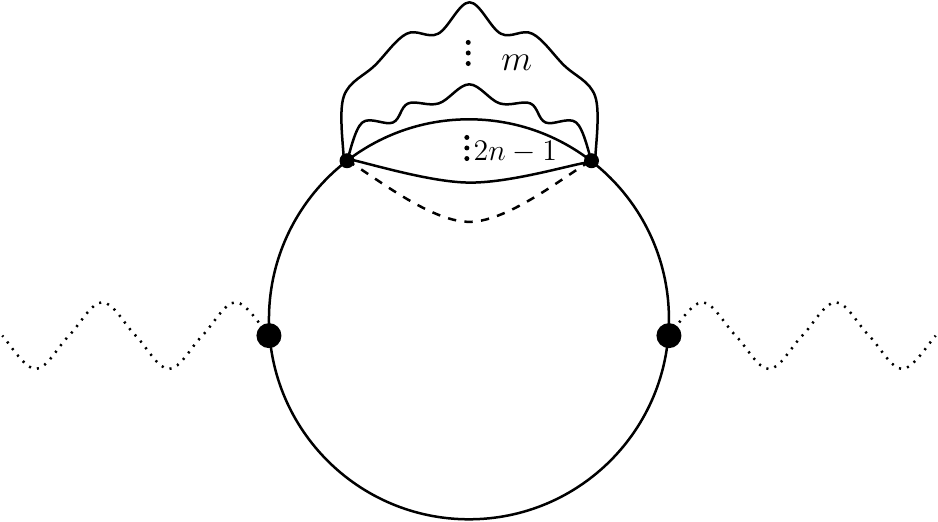}
		\caption{Self-energy correction.}
		\label{fig:jjse}
	\end{subfigure}
		\begin{subfigure}[b]{0.5\columnwidth}
		\centering
		\includegraphics[width=0.9\textwidth]{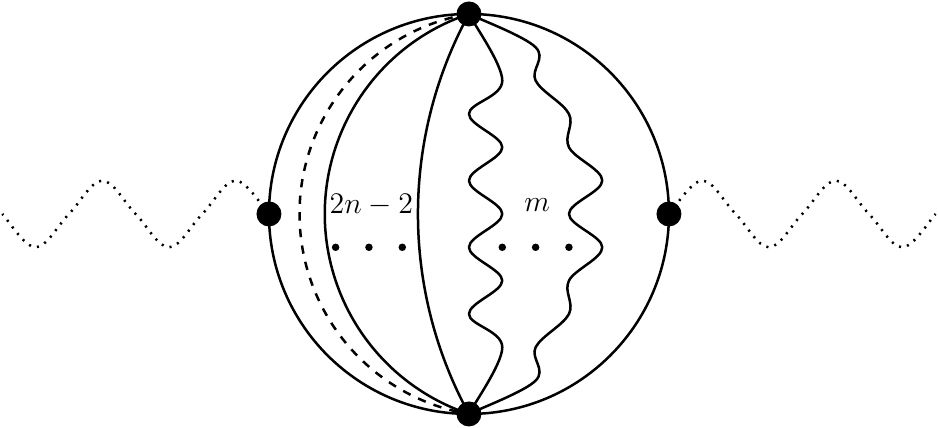}
		\caption{\label{fig:mt}MT diagram}
	\end{subfigure}
	\begin{subfigure}[b]{0.5\columnwidth}
		\centering
		\includegraphics[width=\textwidth]{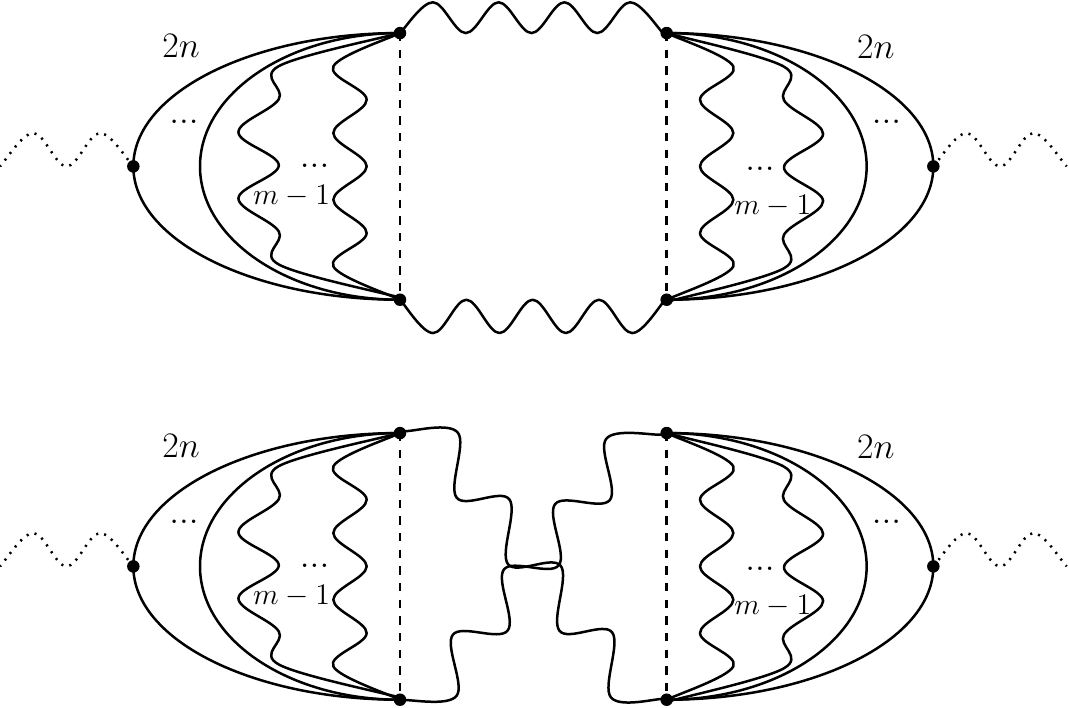}
		\caption{\label{fig:al1}AL diagrams}
	\end{subfigure}
	\caption{\label{fig:polar} Polarisation diagrams. The solid line represents the ``bare'' propagator \eqref{eqn:g1st}.}
\end{figure}

The polarisation bubble receives contributions from the simplest one-loop diagram, as well as from self-energy insertions and vertex corrections, as illustrated in Fig.~\ref{fig:polar}. To leading order, the polarisation is illustrated in Fig.~\ref{fig:jj}, and is given by \cite{Coleman2019}
\begin{eqnarray}
	&&\tilde{\Pi}^{\mu\nu}_0(\bi\Omega_m)\nn
	&=&
	-\frac{2}{m^2}T\sum_{n}\int\frac{d^d\bk}{(2\pi)^2}k^{\mu}k^{\nu}
	\frac{1}{\bi\omega_n-\frac{\bk^2}{2m}+\mu+\bi\Gamma\sgn(\omega_n)/2}
	\frac{1}{\bi(\omega_n+\Omega_m)-\frac{\bk^2}{2m}+\mu+\bi\Gamma\sgn(\omega_n+\Omega_m)/2}\nn
	&\simeq&-v_F^2\mathcal{N}\delta^{\mu\nu}\frac{\Omega_m}{\Omega_m+\sgn(\Omega_m)\Gamma},
\end{eqnarray}
which leads to a finite constant residual resistivity. Here we have assumed that the dominant contributions come from fermions near the Fermi surface \cite{Coleman2019}.\\ 

At the next order, the polarisation receives corrections from the fermionic self-energy, shown in Fig.~\ref{fig:jjse}. For convenience, we write the self-energy in the form
\begin{eqnarray}
	\Sigma(\bi\omega)\equiv -\bi c_F g^2 \omega^{\varsigma}.
\end{eqnarray}
The corresponding correction to the polarisation tensor is
\begin{eqnarray}\label{eqn:polar2}
	\tilde{\Pi}_g^{\mu\nu}(\bi\Omega)
	&\sim&-\frac{2k_F^2\mathcal{N}}{m^2}\delta^{\mu\nu}\int\frac{d\omega_n}{2\pi}d\xi_{\bk}
	\frac{1}{\bi\omega_n-\xi_{\bk}+\bi\Gamma\sgn(\omega)/2}\frac{1}{\bi\omega_n-\xi_{\bk}+\bi\Gamma\sgn(\omega)/2}\nn
	&&\times 	\frac{1}{\bi(\omega+\Omega)-\xi_{\bk}+\bi\Gamma\sgn(\omega+\Omega)/2} \left(-\bi c_Fg^2\omega^{\varsigma}\right)\nn
	&\sim&g^2\delta^{\mu\nu}\Omega^{\varsigma+1}
\end{eqnarray}
at low frequencies. \\

The vertex corrections, the Maki-Thompson (MT) diagram (Fig.\ref{fig:mt}) and the Aslamazov-Larkin (AL) diagrams (Fig.\ref{fig:al1}), vanish in our model. The disorder average enforces spatial locality, effectively decoupling all momenta in the internal propagators via a delta function in position space \cite{Patel2023}. Take the MT diagram as an example: after disorder averaging, it yields
\begin{eqnarray}
	\tilde{\Pi}_{\text{MT}}(\bi\Omega)\sim\int d^2\bk d^2\bk' \bk\bk'G(\bk,\bi\omega)G(\bk,\bi(\omega+\Omega))G(\bk',\bi\omega')G(\bk',\bi(\omega'+\Omega))...,
\end{eqnarray}
where the integrand is odd under both $\bk\to -\bk$ and $\bk'\to -\bk'$, and thus integrates to zero. The same argument applies to the AL diagrams. As a result, all vertex corrections vanish in these SYK-rised models.\\

Having computed all relevant polarisation tensors, we now apply the Kubo formula to extract the resistivity. Denoting the self-energy correction to the polarisation as $\tilde{\Pi}_g\equiv -c_g g^2\Omega^{1+\varsigma}$, where $c_g$ is a constant, the total polarisation tensor to order $\mathcal{O}(g^2)$ reads
\begin{eqnarray}
	\tilde{\Pi}^{\mu\nu}\simeq \tilde{\Pi}^{\mu\nu}_0+\tilde{\Pi}^{\mu\nu}_g.
\end{eqnarray}
Substituting into the Kubo formula \eqref{eq:kubo} and analytically continuing ($\bi\Omega\to\Omega$), we obtain the conductivity
\begin{eqnarray}\label{eqn:conductivity}
	\sigma^{\mu\nu}(\Omega)=v_F^2\mathcal{N}\delta^{\mu\nu}\frac{1}{\Gamma-\bi\Omega}-c_g g^2\delta^{\mu\nu}(-\bi\Omega)^{\varsigma}.
\end{eqnarray}
In the absence of a magnetic field, the tensor structure is trivial and $\tilde{\Pi}$ is diagonal, so we drop the indices and write the scalar conductivity. Taking the inverse and extracting the real part, the resistivity is
\begin{eqnarray}\label{eqn:resistivity}
	\Re\rho=\Re\frac{1}{\sigma}\sim \frac{2\Gamma}{\mathcal{N}v_F^2}+g^2 c_g' \Omega^{\varsigma},
\end{eqnarray}
where $c_g'$ absorbs numerical prefactors independent of $g^2$.\\

The frequency dependence at zero temperature translates to $T$-dependence at finite temperature. 
Our goal is to identify all combinations of $d,m,n$ that yield a linear resistivity.
\begin{itemize}
	\item First, for $d=2$, according to eqn.\eqref{eqn:d2sigma}, it is to solve
	\begin{eqnarray}
		\varsigma=2n+m-2=1.
	\end{eqnarray}
	Since $n\geq 1$ and $m\geq 1$, only $n=m=1$ can yield a linear resistivity ($\varsigma=1$), which has been studied in ref.\cite{Patel2023}.
	\item If $d=3$ and $\eta>0$, eqn.\eqref{eqn:d3sigma1} requires
	\begin{eqnarray}
		\varsigma=\Theta(m)\frac{\eta}{2}+2n+m-2=1.
	\end{eqnarray}
	As $m\geq 1$, $\Theta(m)\eta>0$, while $2n+m-2\geq 1$. So this equation has no solution. That is, there is no $\{n,m,d\}$ can yield linear resistivity in this case.
	\item If $d=4$ and $\eta>0$, eqn.\eqref{eqn:d3sigma1} requires
	\begin{eqnarray}
		\varsigma=\Theta(m)\eta+2n+m-2=1.
	\end{eqnarray}
	Similarly, there is no solution.
	\item If $d\geq3$ and $\eta<0$, eqns.\eqref{eqn:d3sigma2} and \eqref{eqn:d4sigma2} implies
		\begin{eqnarray}\label{eqn:varsigma2}
		\varsigma=\Theta(m)\frac{\eta m (d-2)}{2}+2n+m-2=1.
	\end{eqnarray}
	Since $\eta<0<2$, this also requires
	\begin{eqnarray}
		\eta=\Theta(m-1)\frac{\eta(m-1)(d-2)}{2}+2n+m-2.
	\end{eqnarray}
	When $m=1$, $\eta=2n+m-2>0$, contradicting with the condition that $\eta<0$, so we should assume $m\geq 2$. One finds
	\begin{eqnarray}\label{eqn:eta}
		\eta=-\frac{2 (m+2 n-2)}{d m-d-2 m}<0.
	\end{eqnarray}
	Substituting to eqn.\eqref{eqn:varsigma2}, one obtains
	\begin{eqnarray}
		\varsigma=-\frac{d (m+2 n-2)}{d (m-1)-2 m}=1,
	\end{eqnarray}
	which can be rearranged as
	\begin{eqnarray}\label{eqn:n}
		n=\left(\frac{1}{d}-1\right) m+\frac{3}{2}\in\mathbb{Z}^+.
	\end{eqnarray}
	Since $n\geq1$, $m$ should satisfies
	\begin{eqnarray}
		0<m\leq \frac{1}{2}\frac{d}{d-1}.
	\end{eqnarray}
	The only integer $m$ satisfying this is $m=1$ when $d=2$. This contradicts with the requirement that $m\geq 2$ and $d\geq 3$. So there is no solution for $\{n,m,d\}$ giving rise to linear-$T$ resistivity in this case either.
\end{itemize}

In summary, \emph{the only consistent choice yielding linear resistivity is $d=2$ with $n=m=1$}, precisely corresponding to the Yukawa-type coupling studied in ref.~\cite{Patel2023}. Consequently, the potential divergence in eqn.\eqref{eqn:dintegral1} discussed above does not exist in this case, since it is irrelevant to control the divergences in other models that does not give rise to linearity. \\

As mentioned before, we ignore potential divergence and focus primarily on the scaling behaviour. Given that $m=n=1$ and $d=2$ is the only solution to $\varsigma=1$, let us neglect other models in this family, and briefly discuss the divergence in this theory.
When one performs explicit computation in $2D$ Yukawa-SYK model \eqref{eqn:yukawa}, there will be a UV divergence entering the electron self-energy, which reads \cite{Patel2023}
\begin{eqnarray}
	\Sigma\sim \omega_n\ln(\frac{\me\Lambda_q^2}{|\omega_n|}),
\end{eqnarray}
where $\Lambda_q$ is the UV cutoff on boson momentum $\bq$. Such a divergence will enter the polarisation, such that
\begin{eqnarray}
	\sigma(\bi\Omega_m)\sim \Omega_m\ln(\frac{\me^3 \Lambda_q^4}{\omega_n^2}).
\end{eqnarray} 
As we will only take the \emph{real} part of the conductivity, after an analytical continuation ($\bi\Omega_m\to-\Omega+\bi0^+$), one obtains
\begin{eqnarray}
	\sigma(\Omega)&\sim&\bi\Omega\left(\ln(-\me^3\Lambda_q^4)-\ln(\Omega^2)\right)\nn
	&=&\bi\Omega\left(\ln(-1)+\ln(\me^3\Lambda_q^4)-\ln(\omega^2)\right)\nn
	&=&\Omega\left(-\pi+\bi\ln(\me^3\Lambda_q^4)-\bi\ln(\omega^2)\right).
\end{eqnarray}
Therefore, the real part of the conductivity is linear in frequency, whilst the divergence only appears in the imaginary part, which is of no importance in this case. Therefore, neither of these two points affects our main conclusion in this paper.

\section{Spatially Uniform Model}\label{sec:uniform}
This section will offers a heuristic argument for that sptatially uniform coupling between FS and critical bosons can never yield linear resistivity, based on dimensional analysis. To be precise, we consider the same model, while the coupling parameter has no spatial dependence. That is,
\begin{eqnarray}\label{eqn:su}
	S_{g}
	&=&\frac{g_{\{i\}\{j\}\{l\}}}{N^{(2n+m-1)/2}} \int d\tau d^d r \sum_{\{i\},\{j\},\{l\}}
	\psi^\dagger_{i_1,\bk}(\br,\tau)...\psi^\dagger_{i_n,\bk}(\br,\tau)\nn
	&\times&\psi_{j_1,\bk+\bq}(\br,\tau)...\psi_{j_n,\bk+\bq}(\br,\tau)\phi_{l_1,\bq}(\br,\tau)...\phi_{l_m,\bq}(\br,\tau),
	\label{eq:action2}
\end{eqnarray}
with
\begin{eqnarray}
	&&\langle g_{\{i\}\{j\}\{l\}}\rangle=0,\\
	&&\langle g_{\{i\}\{j\}\{l\}}g^*_{\{i'\}\{j'\}\{l'\}}\rangle=g^2\delta_{\{i\}\{i'\},\{j\}\{j'\},\{l\}\{l'\}}.
\end{eqnarray}
Firstly, without potential disorder, the system shows a vanishing resistivity due to the `boson drag' \cite{Guo2022}, so random couplings \eqref{eqn:su} without potential disorderr have been ruled out at the beginning.\\

Due to momentum conservation,
the integral \eqref{eqn:dintegral1} becomes
\begin{eqnarray}\label{eqn:intloc}
	\mathfrak{I}^d_{\alpha,\beta}(\bi x,\bm y)
	&\equiv&\int_{-\infty}^{\infty}\left(\prod_{i=1}^{\alpha-1}\frac{d\omega_i}{2\pi}\frac{d^d\bk_i}{(2\pi)^d}\frac{1}{\bi\sgn(\omega_i)\Gamma/2-\xi_{\bk_i}}\right)\left(\int_{-\infty}^{\infty}\prod_{j=1}^{\beta}\frac{d\Omega_j}{2\pi}\int_{0}^{\infty}\frac{d^d\bq_j}{(2\pi)^2}
	\frac{1}{c_B|\Omega_j|^\eta+\bq_j^2}\right)\nn
	&&\times
	\frac{1}{\bi\sgn(x+\sum_{j=1}^{\beta}\Omega_j+\sum_{i=1}^{\alpha-1}\omega_i)\Gamma/2-\xi_{\sum_{i=1}^{\alpha-1}\bk_i+\sum_{i=1}^{\beta}\bq_i+\bm y}}.
\end{eqnarray}
The integral $\int d^d\bq$ should be decomposed into $\int d\theta$ and $\int q dq$, and the scattering angles $\theta_i$ will appear in $\xi_{\sum_{i=1}^{\alpha-1}\bk_i+\sum_{i=1}^{\beta}\bq_i+\bm y}$ after expansion.
The total integral is difficult to be computed analytically, so we apply a dimensional analysis.
\begin{itemize}
	\item[1.] One first notices that $x$ is the frequency, so $[x]=[\omega]=[\Omega]$.
	The disorder scattering rate $\Gamma$ is the largest scale in our system, so the angular integral can be neglected, together with the $q$-dependence in these terms. Consequently, $\int d\theta$ brings a term of $[x]^0[q]^0$.
	\item[2.] Each integral $\int d^d \bq$ will result in a polynomial containing terms of dimensions from $[q]^{d-2}[\Omega]^0$ to $[q]^0[\Omega]^{\eta(d-2)/2}$. Imposing a UV cutoff $\Lambda_q$ on $\bq$, we obtain terms from  $[\Omega]^0$ to $[\Omega]^{\beta\eta(d-2)/2}$.
	\item[3.] The integrals $\int d\xi_{\bk}$ are again irrelevant in bringing any scaling behaviour, but they also induces terms $\sgn[\omega_i]$. Because of the symmetry of the integrand, the constant result from $\prod_{j=1}^\beta\int d^d\bq_j$, $\Lambda_q^\beta$ has no contribution. The remaining terms from these integrals are of dimensions from $\Lambda_q^{d-3}[\Omega]^{\eta/2}$ to $\Lambda_q^0[\Omega]^{\eta(d-2)/2}$. If $\eta<0$, the leading term is $\Lambda_q^0[\Omega]^{\beta\eta(d-2)/2}$; if $\eta>0$, $\Lambda_q^{d-3}[\Omega]^{\eta/2}$ is dominant.
	\item[4.] The frequency integrals, $\int d\omega_i$ and $\int d\Omega_j$ together yield a dimension of $[\text{frequency}]^{\alpha+\beta-1}$.  
	\item[5.] Finally, the domain of $\omega_i$ and $\Omega_j$ will be limited within $[0,|x|]$. One finds
	\begin{eqnarray}
		\mathfrak{I}^d_{\alpha,\beta}(\bi x,\bm y)&\sim&x^{\alpha+\beta+\Theta[\beta]\frac{\beta\eta(d-2)}{2}-1},
	\end{eqnarray}
	for $\eta<0$
	and 
	\begin{eqnarray}
		\mathfrak{I}^d_{\alpha,\beta}(\bi x,\bm y)&\sim&x^{\alpha+\beta+\Theta[\beta]\frac{\eta}{2}-1},
	\end{eqnarray}
	for $\eta>0$.
\end{itemize}
This is precisely the same scaling behaviour with $\mathcal{I}^d_{\alpha,\beta}(\bi x,\bm y)$, given in section \ref{sec:model}. \\


So at the level of self-energies, the spatial independent model \eqref{eqn:intloc} has the same scaling behaviour with spatially random model \eqref{eqn:multiple}. According to eqn.\eqref{eqn:conductivity}, the leading-order conductivity from $\tilde{\Pi}_g^{\mu\nu}$ will inherit the freqency-dependence from electron self-energy, so a self-energy that satisfies $\Sigma\sim\omega$ will yield a linear conductivity. However, a spatially uniform system admits vertex corrections, MT diagram and AL diagrams. Therefore, the self-energy alone is not enough the capture the correct transport properties. Suppose $\Sigma\sim\omega^\square$. If $\square<0$, vertex corrections will yield terms $\omega^{\square'}$, with $\square'\leq\square<0$. On the other hand, if $\square>0$, one will find $\square'\geq \square>0$. Therefore, the minimum condition of model \eqref{eqn:intloc} to reproduce a linear resistivity is that $\Sigma\sim \omega^\square$ with $0\leq\square\leq1$.\\

Applying the same analysis in Sec.\ref{sec:conductivity}, we conclude that only $d=2$ and $m=n=1$ can give us electron self-energy linear in frequency, and increasing dimensions or number of the fields will increase the value of $\square$ as well. Under the condition that $\square\geq 0$, $\Sigma\sim\omega$ is the lowest order one can obtain from the interaction \eqref{eqn:su} in $d\geq 2$, and the spatially uniform model with $m=n=1$ in $2$D is the only candidate of strange metals in this family.\\

However, the linear contribution from electron self-energy will be canceled by MT diagrams \cite{Patel2023,Guo2022}, whereas the AL will bring higher order terms. The only candidate is also ruled out, so the interaction \eqref{eqn:su} can never give rise to a linear resistivity.\\

Combining what we have found in sec.\ref{sec:conductivity}, one reaches the conclusion that only $2$D spatially random Yukawa coupling \eqref{eqn:yukawa} is able to reproduce strange-metal transport.

\section{Conclusion}\label{sec:conclusion}
In this work, we have systematically explored the full class of SYK-rised scalar couplings of the form $(\psi^\dagger \psi)^n \phi^m$, involving multiple fermions and scalar bosons, as potential candidate theories for strange metals, across arbitrary dimensions $d\geq2$. 
It turns out that only spatially random Yukawa-type interactions ($n=m=1$) in $(2+1)$ dimensions lead to linear-$T$ resistivity. No other combination within this interaction class leads to such scaling, making the 2D Yukawa model the unique candidate among scalar-coupled SYK-rised theories.\\

In earlier work \cite{Wang:2024utm}, we also demonstrated that a spatially random QED-like vector coupling, $\psi^{\dagger}\nabla^{\mu}\psi\ba_{\mu}$, yields linear resistivity. Here $\ba$ is a bosonic vector field, not necessarily a gauge field. 
Taken together, these results suggest that in realistic two-dimensional systems, the random Yukawa and QED-like couplings serve as minimal building blocks for capturing linear-$T$ resistivity. In $(2+1)$D systems, any interaction of the form \eqref{eqn:multiple} results in a low-temperature resistivity scaling as $\rho \sim T^{m + 2n - 2}$. Thus, the linear case arises only when $n = m = 1$, or equivalently from the vector coupling scenario. All higher order terms do not contribute to the linear resistivity. Therefore, there are essentially one class in each category: unique scalar class and unique vector class. \\

In addition to the classification itself, the results suggest a more general lesson. Spatial randomness alone is insufficient to produce strange-metal behaviour. The transport properties are strongly constrained by both the structure of the interaction and the spatial dimension. Within the SYK-rised scalar couplings $(\psi^\dagger\psi)^n\phi^m)$, increasing the number of fields drives the resistivity away from linearity, while changing the spatial dimension modifies the scaling relations too. As a consequence, the majority of candidate interactions are excluded, leaving only the two-dimensional Yukawa coupling as a viable source of linear-$T$ resistivity. In this sense, the present work is not merely an identification of a successful model, but also a demonstration of how narrowly constrained strange-metal behaviour is within this class of theories. The distinguished role of two dimensions is also noteworthy. Within the SYK-rised framework considered in this article, the ubiquity of strange-metal phenomenology in quasi-two-dimensional materials may reflect a deeper connection between dimensionality and transport scaling.\\

This paper has focused exclusively on low-temperature linear resistivity in SYK-rised models. While the Yukawa coupling emerges as a viable theory of strange metals within this regime, a more complete understanding requires going beyond the infrared limit. 
It remains an open question whether SYK-rised models can also account for $H$-linear resistivity and magnetoresistance obeying $H/T$ scaling \cite{Jiang2023}. These directions will be crucial for determining whether SYK-rised interactions offer a unified framework for understanding the strange metal phase.

\acknowledgments
The authors would like to thank Ki-Seok Kim, Kyoung-Min Kim, Chao-Jung Lee, Sung-Sik Lee, Yi Zhang for the inspiring discussion. This work is supported by the National Research Foundation of Korea (NRF) grant funded by the Korean government (MSIT) No.RS-2026-25472139. YLW is supported by an appointment to the Young Scientist Training Programme at the APCTP through the Science and Technology Promotion Fund and Lottery Fund of the Korean Government.

\appendix
\section{Useful Integrals}\label{app:math}
The following integrals are useful in evaluating self-energies in this article.
\begin{itemize}
	\item
	\begin{eqnarray}
		\int \frac{x^{d-1}}{A+x^2} dx=\frac{x^d \, _2F_1\left(1,\frac{d}{2};\frac{d}{2}+1;-\frac{x^2}{A}\right)}{A d},
	\end{eqnarray}
	where $_2F_1(a,b;c;d)$ is the hypergeometric function.
	\item
	\begin{eqnarray}\label{eq:j}
		&&\int \left(\prod_{i=1}^{n}dx_i\sgn(x_i)\right)\Big|B-\sum_{i}^{n}x_i\Big|^{\lambda}\nn
		&\sim&\underset{|\sum_{i}^{n}x_i|\leq |B|}{\int_0^{\infty}} \prod_{i=1}^{n}dx_i\Big|B-\sum_{i}^{n}x_i\Big|^{\lambda}
		=\underset{|\sum_{i}^{n}x_i|\leq |B|}{\int_0^{\infty}} \prod_{i=1}^{n}dx_i\Big|B-\sum_{i}^{n}x_i\Big|^{\lambda}\int dX\delta(X-\sum_{i}^{n}x_i)\nn
		&\sim&\int_0^{|B|}d|X|\prod_{i=1}^{n-1}\int_{0}^{|X|}dx_i\Big|B-X\Big|^{\lambda}
		=\int_{0}^B dX |X|^{n-1}\Big|B-X\Big|^{\lambda}\nn&\sim& |B|^{n+\lambda},
	\end{eqnarray}
\end{itemize}

\section{A detour: the failure of Fermi's golden rule}
Beyond resistivity scaling, our study also uncovers the breakdown of Fermi’s Golden Rule in disordered models, even in cases where quasiparticles remain well-defined. \\

Let us detour to the failure of Fermi's golden rule in SYK-rised models.
According to the analysis in ref.\cite{Wang:2024utm}, Fermi's Golden Rule can help to understand the emergence of linearity from the spatial random coupling \eqref{eqn:yukawa}, depite the absence of well-defined quasiparticles. In both scalar and vector models, the bosonic self-energy takes the form $\Pi(\Omega)\sim C_b \Omega$,
where $C_d$ is a constant,
leading to a bosonic dispersion $\Omega\sim \bq^2$. In  $(d+1)$-dimensional systems, Bosonic density of states contributes a $T^{d/2}$-dependence to the resistivity. Meanwhile, the angular correction factor $(1-\cos\theta)$, with $\theta$ the scattering angle, is temperature-independent due to the relaxation of momentum conservation at each interaction vertex. Consequently, at low temperature, both interactions produce linear resistivity, though the linearity arises from different polarisation bubbles in Kubo formula \cite{Wang:2024utm}. \\

In our generalised model \eqref{eqn:multiple}, the bosonic dispersion relation takes the form
\begin{eqnarray}\label{eqn:dispersion}
	\Omega\sim \bq^{2/\eta}\equiv\bq^\alpha,
\end{eqnarray}
where $\alpha$ can be derived from the boson propagator. In section \ref{sec:model}, we have obtained $\Pi\sim\Omega^{\eta'}$, where
\begin{eqnarray}
	\eta'=\left\{
	\begin{aligned}
		2n+m-2, &\quad d=2\\
		\Theta(m-1)\frac{\eta}{2}+2n+m-2, &\quad d= 3, \quad \eta>0\\
		\Theta(m-1)\eta+2n+m-2, &\quad d\geq 4, \quad \eta>0\\
        \Theta(m-1)\frac{\eta(m-1)(d-2)}{2}+2n+m-2, &\quad d\geq 3, \quad \eta<0		
	  	\end{aligned}
	\right.
\end{eqnarray}
In order to determine $\alpha$, we need to compare $\eta'$ with 2. If $\eta'>2$, we have $\alpha=1$; otherwise, $\eta=\eta'$ and $\alpha=2/\eta'$.\\

Here is a quick analysis. 
\begin{itemize}
	\item[i)]  We can see that when $m=n=1$, $\eta=\eta'=1$ in all dimensions $d\geq 2$, so $\alpha=2$. \footnote{When $\eta<0$, we have $\eta'=\eta<0$, so it is necessary to require $m\geq 2$ in this case. Therefore, the cases where $m=1$ can only be true when $d=2$ or $\eta>0$ in higher dimensions.}
	\item[ii)] $m\geq 2$ or $n\geq 2$ and $d=2$, we always have $\eta'\geq 2$, so $\eta=2$ and $\alpha=1$.
	\item[iii)] If  $m\geq 2$ or $n\geq 2$ in $d\geq 3$ with $\eta>0$, we find $2n+m-2+\eta>2n+m-2+\eta/2>2n+m-2\geq 2$, so $\eta=2$ and $\alpha=1$ too.
	\item[iv)] For $d\geq 3$, $\eta<0$, and $m\geq 2$, we should require $\eta=\eta'$ to make $\eta<0$ possible. This leads to
	\begin{eqnarray}
		\eta=-\frac{2 (m+2 n-2)}{d (m-1)-2 m},
	\end{eqnarray}
	and 
	\begin{eqnarray}
		\alpha=-\frac{d (m-1)-2 m}{m+2 n-2}.
	\end{eqnarray}
\end{itemize}

In summary,
\begin{eqnarray}\label{eqn:alpha}
	\alpha=\left\{
	\begin{aligned}
	    2-\Theta(m+n-2), &\quad d=2\\
	    2-\Theta(m+n-2), &\quad d\geq3,\quad \eta>0\\
		-\frac{d (m-1)-2 m}{m+2 n-2}, &\quad d\geq 3,\quad \eta<0\\
	\end{aligned}\right.
\end{eqnarray}
as implied by the bosonic dispersion from eqns. \eqref{eqn:d2pi} - \eqref{eqn:d3sigma2}.\\

\paragraph{Golden Rule estimate} According to Fermi’s golden rule, the lifetime of an electron is estimated as
\begin{eqnarray}
	\frac{1}{\tau} &\sim& \int \prod_{i=1}^{2n-1} d^d\bk_i\, f_{\bk_i} \prod_{j=1}^{m} d^d\bq_j\, b_{\bq_j} (1 - \cos\theta)\, \delta\left(\sum_{i=1}^{2n-1}\omega_i + \sum_{j=1}^{m} \Omega_j - \omega_{2n}\right)\nn
	&\sim& \int \prod_{i=1}^{2n-1} d\omega_i\, \frac{1}{e^{\beta\omega_i}+1} \prod_{j=1}^{m} d\Omega_j\, \frac{1}{e^{\beta\Omega_j}-1} \Omega_j^{\frac{d}{\alpha} - 1} (1 - \cos\theta)\, \delta\left(\sum_{i=1}^{2n-1}\omega_i + \sum_{j=1}^{m} \Omega_j - \omega_{2n}\right),\nn
\end{eqnarray}
where $f_{\bk}$ and $b_{\bk}$ are the Fermi-Dirac and Bose-Einstein distribution functions. In second step, we have changed variables from momenta to frequencies using despersion relation \eqref{eqn:dispersion}. 
Rescaling variables as $x_i\equiv\beta\omega_i$, $y_j\equiv\beta\Omega_j$, and $z=\beta\omega_{2n-1}$, we find
\begin{eqnarray}\label{eqn:gr}
	\frac{1}{\tau}_{\text{GR}} &\sim& T^{m \frac{d}{\alpha} + 2n - 2} \int \prod_{i=1}^{2n-2} dx_i\, \frac{1}{e^{x_i}+1} \prod_{j=1}^{m} dy_j\, \frac{1}{e^{y_j}-1} \frac{1}{e^z + 1}(1 - \cos\theta).
\end{eqnarray}
\\

\paragraph{Kubo Formula calculation} Meanwhile, the resistivity \eqref{eqn:resistivity} derived from the Kubo formula scales as 
\begin{eqnarray}
	\rho_{\text{Kubo}}\sim T^{\varsigma},
\end{eqnarray}
where 
\begin{eqnarray}\label{eqn:varsigma}
	\varsigma=\left\{
	\begin{aligned}
		2n+m-2, &\quad d=2\\
		\frac{1+\Theta(m+n-2)}{2}+2n+m-2, &\quad d\geq3, \quad \eta>0\\
		\Theta(m+n-2)+2n+m-1, &\quad d\geq4, \quad \eta>0\\
		-\frac{d (m+2 n-2)}{d (m-1)-2 m}=\frac{d}{\alpha}, &\quad \quad d\geq3, \quad \eta<0
	\end{aligned}\right. .
\end{eqnarray}

We see that the golden rule fails to reproduce the correct scaling: the result \eqref{eqn:gr} does not match the Kubo-derived scaling \eqref{eqn:varsigma},  except in the special case $m=n=1$ in $d=2$ and $d=3$,  where the two estimates coincide \cite{Wang:2024utm}. Remarkably, this agreement occurs despite the absence of well-defined quasiparticles in two dimensions.\\

As discussed in the previous section, momentum conservation is relaxed at each interaction vertex due to disorder averaging. This decouples spatial momentum across different interaction lines. It is therefore highly likely that the golden rule, which relies on well-defined scattering kinematics, fails to correctly estimate the resistivity in SYK-rised models, even in the presence of quasiparticles. 
The mechanisms governing when the golden rule applies or fails in SYK-rised models remain unknown and deserve further investigation.


\bibliographystyle{JHEP}
\bibliography{biblio.bib}

\end{document}